\documentclass[proceedings,onecolumn]{rmaa}
\usepackage{rmaacite}

\newcommand{\kms}{\,\mbox{km s$^{-1}$}}

\renewcommand{\P}[1]{%
\ifnum#1=1\hbox{OW~168--326E}\fi \ifnum#1=2\hbox{OW~167--317}\fi
\ifnum#1=3\hbox{OW~163--317}\fi \ifnum#1=5\hbox{OW~158--323}\fi
\ifnum#1=0\hbox{OW~171--334}\fi}

\title{The Broad Line Region in Active Galactic Nuclei}
\author{D. Dultzin-Hacyan\altaffilmark{1},
P. Marziani\altaffilmark{2},  J.W. Sulentic\altaffilmark{3}} \altaffiltext{1}{Instituto
de Astronom\'{\i}a, UNAM, M\'exico} \altaffiltext{2}{Osservatorio Astronomico, Padova,
Italia} \altaffiltext{3}{Dept. Physics and Astronomy, University of Alabama,
Tuscaloosa, USA}

\shortauthor{Dultzin-Hacyan et al.} \shorttitle{BLR in AGN}

\keywords{galaxies: active ---  lines: profiles --- quasars: emission lines}

\abstract{We review constraints on models of the broad line region that are imposed by
observations of the emission lines in Active Galactic Nuclei.  Comparison of  high and
low ionization lines in sources with FWHM H$\beta\leq$4000 km s$^{-1}$ points toward low
ionization line emission produced in a flattened geometry (the accretion disk?) with an
associated  high ionization wind. It remains unclear whether these results can be
extended to all radio quiet AGN and particularly to radio loud AGN.}

\resumen{Revisamos las limitaciones que se pueden imponer a partir de las
observaciones, sobre los modelos de la estructura de las regiones que emiten las lineas
anchas en los N\'ucleos Activos de Galaxias.  La comparaci\'on entre  las lineas de
baja- y alta ionizaci\'on, sugiere que la emisi\'on proviene, en parte, de una
geometr\'{\i}a fuertemente aplanada (el disco de acreci\'on?) y, en parte, de un viento
asociado, para una fracci\'on importante  de objetos radio callados que incluye a los
nucleos Seyfert 1 de lineas angostas (NLSy1). Sin embargo, es incierto si estos
resultados pueden extenderse a todos los Nucleos Activos radio callados y, sobre todo,
a los radio fuertes.}

\nonstopmode

\def\hbnc{{\sc{H}}$\beta_{\rm NC}$\/}
\def\lya{{\sc L}{\rm y}$\alpha$\/}

\def\mgii{{\rm Mg\sc{ii}}$\lambda$2800\/}

\def\o4363{{\sc{[Oiii]}}$\lambda$4363\/}
\def\oiiiuv{{\sc{Oiii]}}$\lambda$1663\/}
\def\heiiuv{{\rm He\sc{ii}}$\lambda$1640}

\def\feiiuv{\rm Fe{\sc ii}$_{\rm UV}$\/}

\def\feii{\rm Fe{\sc ii}}

\def\feiiopt{\rm Fe{\sc ii}$_{\rm opt}$\/}

\def\heii{He{\sc{ii}}$\lambda$4686\/}
\def\oiii{{\sc [Oiii]}$\lambda\lambda$4959,5007}

\def\apj{{\it Ap. J.}}
\def\aap{{\it Astr. Astroph.}}
\def\apjs{{\em Ap. J. Suppl.}}

\def\kms{km~s$^{-1}$}
\def\l{$\lambda$}

\def\mnras{{\it MNRAS}}
\def\aap{{\it Astron. Astrophys.}}
\def\apjl{{\it Ap. J. Lett. }}

\def\ltsima{$\; \buildrel < \over \sim \;$}
\def\simlt{\lower.5ex\hbox{\ltsima}}
\def\gtsima{$\; \buildrel > \over \sim \;$}
\def\rfe{R$_{\rm FeII}$}

\def\gs{$\Gamma_{\rm soft}$}

\def\simgt{\lower.5ex\hbox{\gtsima}}            

\def\hi{{\sc Hi\/}}

\def\ha{{\sc H}$\alpha$}
\def\nv{{\sc Nv}$\lambda$1240}
\def\civ{{\sc{Civ}}$\lambda$1549\/}
\def\civnc{{\sc{Civ}}$\lambda$1549$_{\rm NC}$\/}
\def\civbc{{\sc{Civ}}$\lambda$1549$_{\rm BC}$\/}

\def\cm3{cm$^{-3}$\/}

\def\hb{{\sc{H}}$\beta$\/}

\def\hbbc{{\sc{H}}$\beta_{\rm BC}$\/}
\begin{document}
\maketitle
\section{Introduction \label{sec:intro}}

The Broad Line Region (BLR) in Active Galactic Nuclei (AGN) is unresolved with present
day imaging detectors and it will remain so for the foreseeable future. This is why
``one quasar spectrum is really worth a thousand images'' as stressed  by Gary Ferland
at this meeting. In response we would add that {\em a thousand spectra  are better than
one average spectrum}. Understanding the diversity in optical spectroscopic properties
of AGN  is the key to any realistic AGN modeling  \cite{bg92,smd}.  The ideal to
reconstruct the BLR velocity field from a single profile is not realistic \cite{pen90}.

Determination of BLR structure and kinematics can be approached in two ways. It has
been recognized for a long time that strong broad and narrow emission lines coming from
both high and low ionization species are present in Seyfert galaxies and quasars. This
is considered a defining spectroscopic property of AGN.  Restricting attention to broad
lines: a) typical (i.e., strongest and most frequently observed) high ionization lines
(HIL: ionization potential $\simgt$ 50 eV) are \civ\ \heii\ and \heiiuv\ lines; while
b) observed low ionization lines (LIL:  ionization potential $\simlt $ 20 eV) include
\hi\ Balmer lines, \feii\ multiplets, \mgii, and the Ca{\sc ii}\ IR triplet.

The  first approach involves the study of line variability in response to continuum
changes. This approach  has been pursued through a number of successful monitoring
campaigns using Reverberation Mapping (RM) techniques.  RM requires a large amount of
telescope time and, consequently, has been achieved only for a handful of sources. RM
confirms that photoionization is the main heating process in the BLR and that a large
part of the BLR is optically thick to the ionizing continuum (e. g. \pcite{baldwin97}).
RM studies have quantified a main difference between HIL and LIL; HIL respond to
continuum changes with a time delay of a few days while the LIL respond with a delay of
tens of days. This implies  that the LIL are emitted at larger distance from the
continuum source (\pcite{ulr,kor95}.  An exhaustive list of reference can be found in
\pcite{smd}). RM applied to line profiles suffers from uncertainty in our knowledge
about the physics of continuum and broad line formation, so that conflicting models can
still describe the same lag times \cite{pen91,wanpet96}.

The second approach involves statistical analysis of large samples of line profiles
which differ because of properties that may affect the BLR structure,  for example
samples of  radio quiet (RQ) and radio loud (RL) AGN. Statistical studies can be done
for a single line or by comparing lines sensible to different physical parameters (e.g.,
strongest LIL and HIL). The statistical  approach, on which we will focus here, is more
empirical and therefore requires a conceptual framework for interpretation. This
approach is often criticized as relying on several assumptions including that the
profile variability does not influence profile shapes and that the non-simultaneity of
the observations of LIL and HIL are unimportant. Actually as the size of high quality
data samples grow these effects become less and less important.

\section{Baring the Broad Profiles \label{bare}}

The collection of moderate  resolution  (\l/$\Delta$\l $\sim 10^3$) optical and UV
spectra of good quality (S/N $\simgt$ 20 in the continuum) has become possible only in
recent years thanks to the widespread use of CCD detectors and the unprecedented
sensitivity and resolution of the UV spectrographs on board HST. For practical
purposes, \hi\ \hb\ and \civ  can be considered representative of HIL and LIL
respectively. They are also the best lines for statistical studies because they permit
comparison in the same sources out to z=1.0.  The linearity of response of the
currently employed detectors  has made possible a reliable correction for emission
features contaminating \hb\ and \civ, which are: (1) \feii\ emission; significant
\feiiuv\ emission contaminates the red wing of \civ, and has been identified in I Zw 1
by \pcite{m96} and later confirmed by \pcite{lao97}; (2) \heii\ emission for \hb, and
\heiiuv\ + \oiiiuv\ for \civ; (3) \oiii\ for \hb; (4) narrow component, present in
several cases in both \hb\ and \civ.

\civ\ often shows a narrow core with FWHM $\sim$ 1000-2000 \kms, which is
systematically broader than the narrow component of \hb\ (\hbnc). The separation
between the broad component \civbc\ and the core component is often  ambiguous. This
core however shows no shift with respect to the AGN rest frame \cite{brot94}, no
variations \cite{turla}, and correlates with \oiii\ prominence \cite{franc92}. It can
therefore be ascribed to the NLR and considered as the narrow line component of \civ\
(\civnc). The different width between \civnc\ and \hbbc\ can be understood in terms of
density stratification \cite{sm99} without invoking an additional emitting region such
as the so-called ``intermediate line region''. Even if \civnc\ fractional intensity is
small ($\sim$ 10\%), and in some cases obviously absent, failure to account for \civnc\
has led to the erroneous conclusions that FWHM \civbc\ $>$ FWHM \hbbc\ and that the
\civ\ peak shows no shift with respect to \hb\ \cite{cb96}.

\section{A DIFFERENT BLR STRUCTURE IN RADIO LOUD AND RADIO QUIET AGN? \label{struct}}

\begin{figure}
    \leavevmode
    \includegraphics[width=\columnwidth, height=8cm]{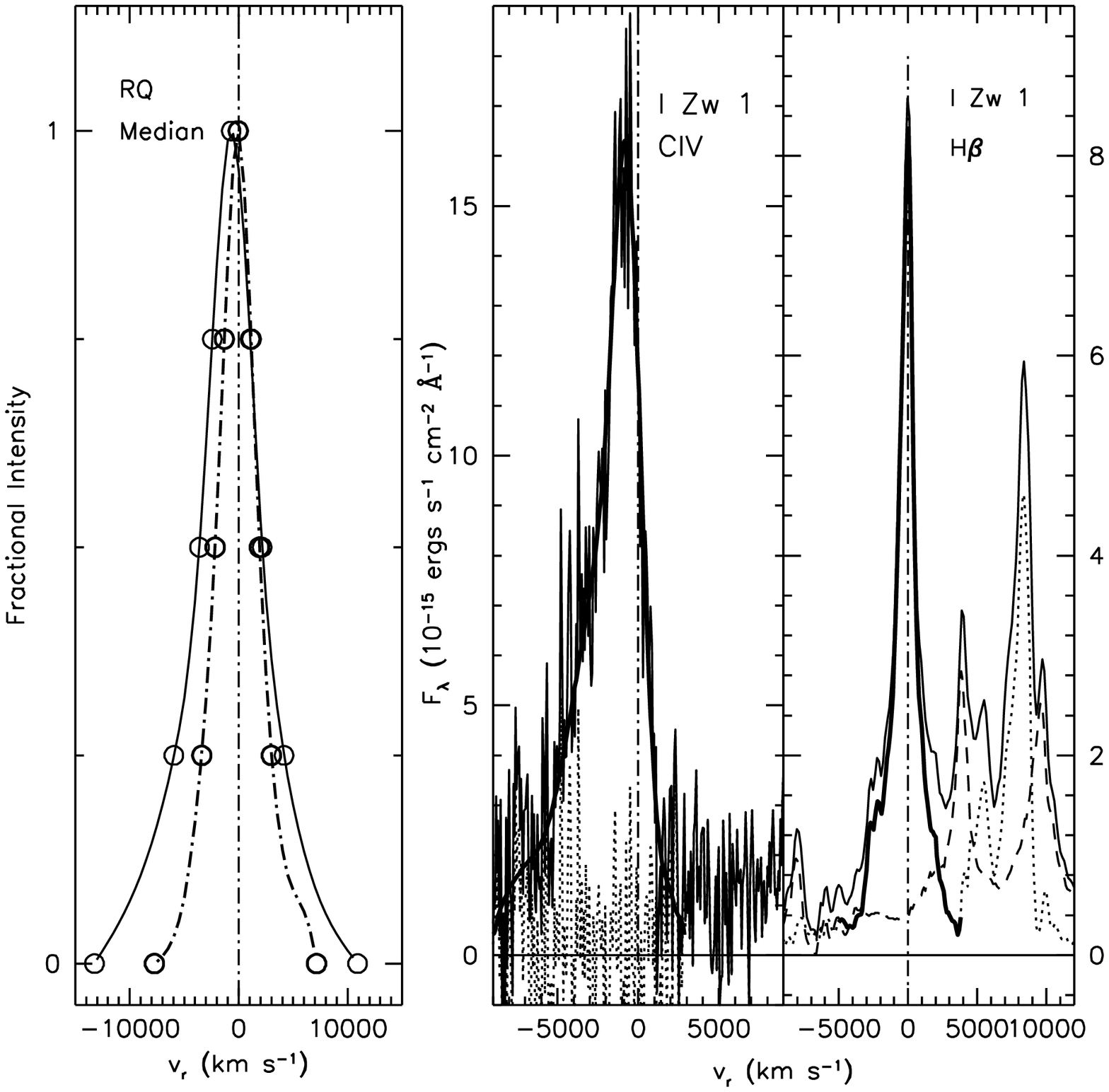}
\caption[]{Profiles of \civbc\ (solid lines) and \hbbc\ normalized to the same peak
intensity and constructed from median values of the radial velocities measured on the
blue and red sides of the profiles for the 21 RQ AGN of \scite{m96} (left panel).
Emission line profiles of \civ\ (middle panel) and \hb\ (right panel) of the prototype
NLSy1 galaxy I Zw 1, adapted from \scite{m96}. The thick and the dotted lines trace \hb\
and \oiii\  after \feiiopt\ (dashed line) subtraction.  \label{fig:rq}}
\end{figure}

\begin{figure}
    \leavevmode
    \includegraphics[width=\columnwidth, height=8cm]{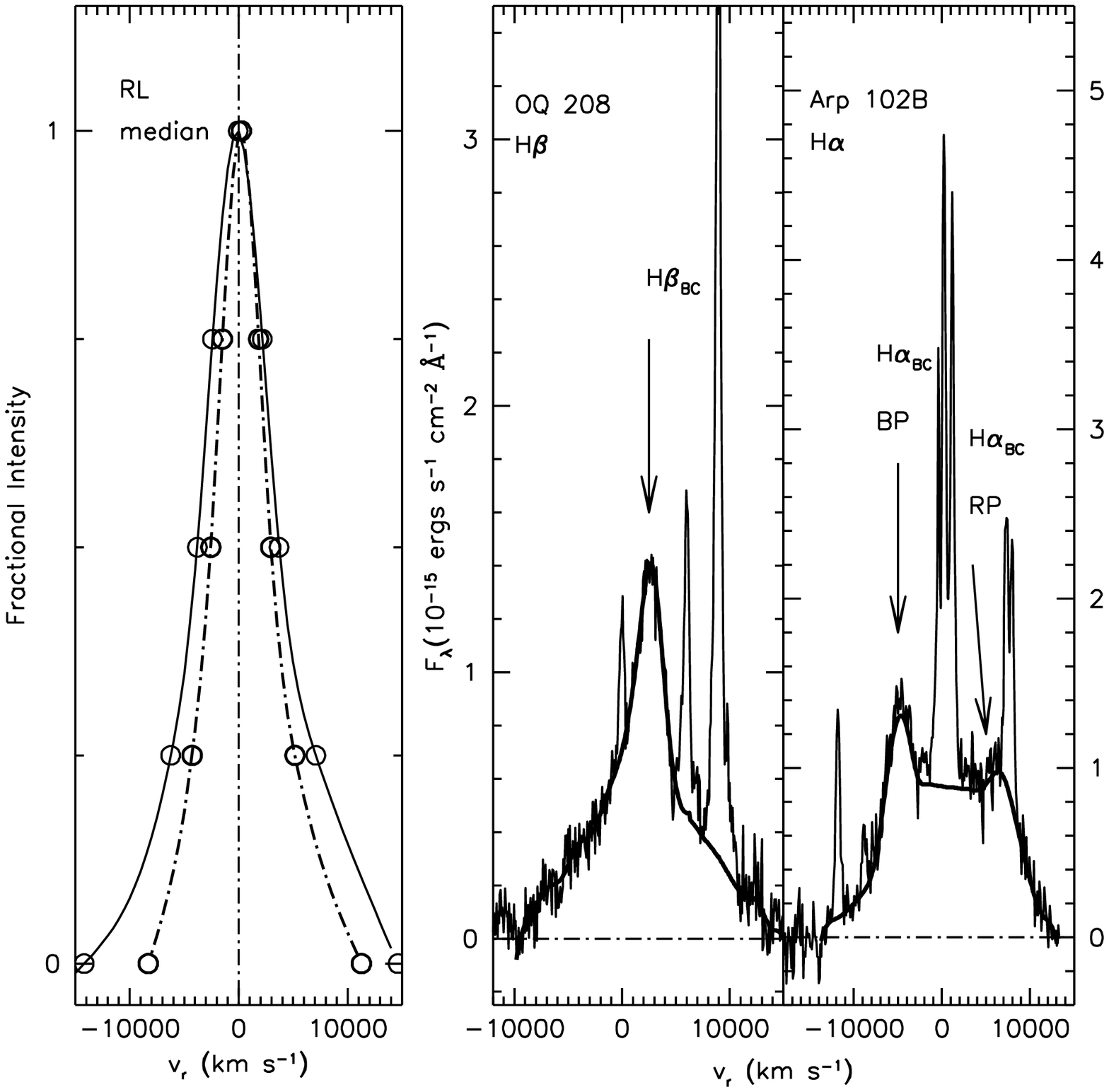}
    \caption[]{Left panel: Profiles of \civbc\ (solid lines) and \hbbc\ normalized to the same
    peak intensity and constructed from median values of the radial velocities measured on
    the blue and red sides of the profiles for the 31 RL AGN of \scite{m96}.
    Middle panel: \hb\ spectrum of OQ 208, showing a single, widely displaced \hbbc\ peak (adapted from \pcite{m93}).
    Right panel: \ha\ profile of the prototype of ``wide-separation double peakers'' Arp 102B.
    Unpublished spectrum obtained at the 1.82 m telescope of the Asiago observatory on March 25, 1989.
    The thick solid line traces \hbbc.}
    \label{fig:rl}
\end{figure}

\scite{m96} made a comparison between \civbc\ and \hbbc\ for a sample of 52 AGN (31
RL).  They presented measures of radial velocity for the blue and red sides of \hbbc\
and \civbc\ at 5 different values of fractional intensity which provide a quantitative
description of the profiles. The reference frame was set by the measured velocity of
\oiii$\lambda$5007$\rm{\AA}$ (IZw1 was the only exception). Standard profile parameters
like peak shift, FWHM, asymmetry index and curtosis can be extracted from these
measures. Representative profiles constructed from the {\em median values} of \civbc\
and \hbbc\ $v_r$\ are reproduced in the left panels of Fig. \ref{fig:rq} and
\ref{fig:rl} for the radio quiet (RQ) and radio loud (RL) samples respectively.

\civbc\ is broader than \hbbc in both RQ and RL samples and it is {\em almost always}
blueshifted relative to \hbbc. However Fig. \ref{fig:rq} and Fig. \ref{fig:rl} show
significant differences between RQ and RL AGN. In RQ AGN, \civbc\ is significantly
blueshifted with respect to the source rest frame while \hbbc\ is symmetric and
unshifted. Contrarily in RL AGN \civbc\ is more symmetric, while \hbbc\ is shifted to
the red at peak intensity and redward asymmetric as well (the median profile
corresponds to the type AR,R according to \pcite{sul89}). There are two important
results which are not displayed in the Figures:  (i) in RQ AGN, \civbc\ blueshifts are
apparently uncorrelated with respect to any \hbbc\ line profile parameter and the
largest \civ\ blueshifts are associated with the lowest W(\civ); (ii) in RL AGN, on the
contrary, \civbc\ and \hbbc\ line profile parameters (FWHM and peak shift) appear to be
correlated. Asymmetry index of \civbc\ and \hbbc, even if not correlated, shows a clear
trend toward asymmetries of the same kind (symmetric or redward asymmetric). These
findings on \civbc\ have been  confirmed by other authors (\pcite{wills95}, save the
difference in terminology and line profile decomposition) and especially by an analysis
of archival HST/FOS observations which have become publicly available after 1995
\cite{sul20}.

The LIL and HIL emitting regions are apparently de-coupled in at least some RQ
sources.  The ``de-coupling'' is well seen in I Zw 1, the prototype Narrow Line Seyfert
1 Galaxy (NLSy1; see Fig. \ref{fig:rq}). The \hb\ profile is very narrow, slightly
blueward asymmetric and unshifted with respect to the rest frame defined by 21 cm
observations, while the \civ\ profile is almost totally blueshifted. At least in the
case of I Zw 1 the distinction between LIL and HIL emitting regions appears to be
observationally established (it was actually suggested because of the difficulty to
explain the relative strengths of LIL and HIL emission using a photoionized ``single
cloud;'' \pcite{cs88}). There is a very important zeroth-order result here: since I Zw
1 is a strong \feiiuv\ emitter, we have HIL \civ\ and LIL \feiiuv\ in the same
rest-frame wavelength range. We see that \feiiuv\ is obviously unshifted (this can be
very well seen by shifting an \feiiuv\ template to the peak radial velocity of \civ).
This result disproves  models that see an (unknown) wavelength dependent mechanism
accounting for the quasar broad line shifts relative to the quasar rest frame.

RL AGN apparently mirror RQ AGN in a curious way: \civbc\ is more symmetric, while
\hbbc\ shows preferentially redshifted profiles and increasingly redward asymmetries.
Large peak redshifts ($v_r \simgt 1000$ \kms, as in the case of OQ 208, Fig.
\ref{fig:rl}) are rarely observed; \hbbc\ peak shifts are usually small ($\Delta
v_r$/FWHM $\ll$ 1, median profile of Fig. \ref{fig:rl}). RL \civbc\ and \hbbc\ data
leave open the possibility that both lines are emitted in the same region.  \civbc\
shows a red-wing (very evident in the latest, higher S/N spectra analyzed by
\scite{sul20}), which cannot be entirely accounted for by \feiiuv emission. It is
interesting to note that superluminal sources with apparent radial velocity $\beta_{\rm
app} \sim 5-10$~ (whose radio axis is probably oriented close to the  line of sight in
the sample of \scite{m96} show very strong \civ\ redward asymmetries, low W(\civ) and
W(\hbbc). This result suggests that redshifts are maximized in RL objects at
``face-on'' orientation (we assume that any disk is perpendicular to the radio axis).

\section{NLSy1 NUCLEI ARE NOT A DISJOINT RQ POPULATION}

NLSy1 are neither peculiar nor rare.  The 8$^{th}$\ edition of the \scite{veron98}
catalogue includes 119 NLSy1 satisfying the defining criterion FWHM Balmer $\simlt$
2000 \kms. They account for $\approx 10$\%\ of all AGN in the same redshift and
absolute magnitude range. Attention toward NLSy1 remained dormant after their
identification as a particular class \cite{op85} until it was discovered that they may
represent $\approx$1/3--1/2 of all soft X-ray selected Seyfert 1 sources (e. g.,
\pcite{grupe98}). NLSy1 are also apparently favored in AGN samples selected on the
basis of color. They account for 27\%\ of the RQ \scite{bg92} sample, probably because
of an optical continuum that is steeply rising toward the near UV.  NLSy1 do not occupy
a disjoint region in parameter  space. They are at an extremum in the FWHM(\hb) vs.
\rfe\ (=I(\feii \l 4570/I(\hbbc)) and in the ``Eigenvector 1'' parameter spaces
\cite{bg92,bf99,smd}. Also, the soft X-ray spectral index \gs\ shows a continuous
distribution which includes NLSy1 at the high end \cite{wang96,smd,sul20}.

Orientation can easily explain much of the RQ  phenomenology observed by \scite{m96}. I
Zw 1 can be considered as an extremum with an accretion disk seen face-on
(i$=0^\circ$), and an outflowing wind observed along the disk axis. We can infer that
the opening angle of any \civ\ outflow is probably large (i. e., the wind is quasi
spherical) because \civ\ profiles like  I Zw 1 are rare. Other NLSy1 show low W(\civ),
and strong \feiiopt, but do not always show large \civ\ blueshifts \cite{rp97}.
Nonetheless,  it is still possible that NLSy1 may be structurally different from other
RQ AGN. If the soft X-ray excess of NLSy1 is due to high accretion rate, then a slim
accretion disk is expected to form  \cite{abram88}. Line correlations presented in
\cite{sul20} appear to hold until FWHM(\hb)$\simlt$ 4000 \kms. For FWHM$\simgt$ 4000
\kms line parameters appear to be uncorrelated however it is still unclear it is at
present unclear because of the difficulty in measuring weak and broad \feiiopt sources
and/or because of a BLR structural difference. This limit may be related to the
possibility of sustaining a particular disk structure and an HIL outflow. A second
parameter, independent from orientation is needed to account for the FWHM(\hbbc) vs
\rfe\ vs \gs\ sequences (see \pcite{smd,sul20} for a detailed discussion).

\section{Inferences on BLR Models for RQ AGN}

Models developed  almost independently of the data through the  80's and early 90's.
The situation has now changed because of three main developments: (1) the ``Eigenvector
1'' correlations allow a systematic view of the change in optical emission line
properties for different classes of AGN \cite{bg92,bf99,smd}; (2) the \civ\ - \hb\
comparison has yielded direct clues about the structure of the BLR \cite{m96,sul20} and
(3) data collected for RM projects provide a high-sampling description of line
variations. For instance, binary black hole scenarios \cite{gask96} were recently
challenged by the failure to detect the radial velocity variations expected from
previous observations and model predictions \cite{eracl97}.

A model in which \civ\ is emitted by outflowing gas (e.g. a spherical wind) while
\hbbc\ is emitted in a flattened distribution of gas (observed in a direction that
minimizes velocity dispersion) such as an optically thick disk (obscuring the receding
half of the \civ\ flow, \pcite{liv97}) or at the wind base  is immediately consistent
with the I Zw 1 data. The big question is whether the results for I Zw 1 can be
straightforwardly extended to other RQ AGN.

An accretion disk (AD) provides a high density and high column density medium for
\feii\ production (e. g. \pcite{dj92}), and possibly other low ionization lines such as
CaII (e. g. \pcite{dd99}). AD avoid conflict with the  stringent restrictions on line
profile smoothness imposed by the first extremely high s/n Balmer line observations
(\pcite{arav97,arav98}) of -incidentally- two NLSy1. They place a lower limit of
(10$^{7-8}$) on the number of discrete emitters needed to explain the observed profiles.

Winds arise as a natural component of an AD model when the effects of radiative
acceleration are properly taken into account \cite{murray95,mc98} or when a
hydromagnetic or hydrodynamic treatment is performed \cite{bottorff97,williams99}.  A
signature of radiative acceleration is provided by observations of ``double troughs'' in
$\approx \frac{1}{5}$ of BAL QSOs i. e., of a hump in the absorption profiles of \nv\
and \civ\  at the radial velocity difference between \lya\ and \nv, 5900 \kms. Such a
feature indicates that \lya\ photons are accelerating the BAL clouds (\pcite{arav94},
and references therein). Additional evidence is provided by the radial velocity
separation in the narrow absortion components of \lya\ and \nv\ which show the same
$\Delta v_r$\ of the two doublet components of \civ\ (e. g. \pcite{wampler91}).

\section{The Trouble With Bare Accretion Disks and Bipolar Flows}

Relativistic Keplerian disks \cite{ch89,sb93} may explain unusual profile shapes (e.g.
double-peaked profiles of Balmer emission lines; \pcite{eh94,sul95}).  Uniform
axisymmetric disk models produce double-peaked line profiles with the blue peak
stronger than the red peak because of Doppler boosting, a feature that is not always
observed in these already rare profiles. To solve this problem, \scite{sb95} and
\scite{eracl95}, proposed that the lines can originate in an eccentric (i.e.
elliptical) disk. Simple disk illumination models can also produce single peaked LIL,
provided they are produced at large radii ($\simgt 10^3$\ gravitational radii) or that
the disk is observed at small inclination \cite{dc90,rokaki92,sul98}. The first of these
conditions may be met in all NLSy1 galaxies; both of them seem to be met in I Zw 1.

Aside from NLSy1 sources, there is general disagreement between observations and model
predictions for externally illuminated Keplerian disks in a line shift-- asymmetry
parameter space \cite{sul90}. Only a minority ($\simlt$ 10 \%) of RL and a handful of
RQ AGN show double peaked Balmer lines suggestive of a Keplerian velocity field
\cite{eh94,sul99}. Double-peakers (e. g. Arp 102B in Fig. \ref{fig:rl},
FWHM(\hbbc)$\simgt$10000 \kms) cannot be like the classical cases because the line
widths are much smaller. The peaks often vary out of phase (Arp 102B: \pcite{mp90}, 3C
390.3: \pcite{zheng91}). Double peaks (NGC 1097) \cite{sb93,sb95} or one of the peaks
(Pictor A) \cite{sul95} sometimes appear quite suddenly. Profile variability studies of
Balmer lines force us to introduce second order modifications to the basic scheme, such
as: eccentric rings and precession \cite{eracl95,sb97}, inhomogeneities such as
orbiting hot spots \cite{zheng91}, and warps \cite{bachev99}. Not even these  {\em
epicycles} are always capable of explaining the observed variability patterns. Even if
elliptical AD models do well in explaining the integrated profiles, they face important
difficulties in explaining variability patterns.

A serious problem for AD models of emission lines is emerging from spectropolarimetric
observations. If  profile broad line shapes are orientation dependent then, in
principle,  the profile shape in polarized light will depend on the distribution of the
scatterers relative to the principal axis and to our line of sight.  Early
spectropolarimetric results showed a discrepancy with the simple disk +
electron-scattering-dominated atmosphere models, which predicted polarization
perpendicular to the radio axis. The observed polarization is low, parallel to the disk
axis, and shows no statistically significant wavelength dependence \cite{anton88}.
Recently \scite{anton96} and \scite{corbett98} included double-peakers in their
samples, and obtained troublesome results for disk emission models because the
polarized \ha\ profiles  are centrally peaked \cite{corbett98}. They investigated the
case of disk emission where the scattering particles are located above and below an
obscuring torus, along its poles. This ``polar scattering model''  is successful in
explaining the polarized profiles {\em  but not the position angle of the polarization
vector}.

The same polarization studies  indicate that the {\em only} scenario that can account
for both the shapes  of  the scattered line profiles and the alignment of the optical
polarization with the radio jet in wide separation double peakers like Arp 102B (Fig.
\ref{fig:rl}) involves a {\em biconical} BLR within an obscuring torus. \ha\ photons
emitted by clouds participating in a biconical flow are scattered towards the observer
by dust or electrons in the inner wall of the surrounding torus. The particular case of
biconical outflow was first developed to reproduce observed profiles  by
\scite{zheng90}. This model has been successfully applied to fit observed profiles in
double-peaked objects \cite{zheng91,sul95}.

Double peaked or single blueshifted peak LIL profiles fitted with bi-cone outflow
models require that the receding part of the flow is also seen. Self-gravity may be
important beyond $\sim 1$ pc, and the disk may be advection-dominated and optically
thin \cite{liv96}. Recent work by \scite{sh99} models the vertical structure of AD and
the origin of thermal winds above AD. They not only find that a wind powered by a
thermal instability develops in all disks with certain opacity laws but also that in
disks dominated by bremsstrahlung radiation, a time-dependent inner hole develops below
a critical accretion rate. This scenario provides a natural explanation for transient
double-peakers, such as NGC1097 \cite{sb95,sb97}, but low accretion rate is a
requirement for {\em both} advection dominated disks and hole formation.

\scite{sul95} explored  the idea that the double-peaked emitters represent a
geometrical extremum where an outflow is viewed close to pole-on. However,
double-peakers are associated with double-lobe radio-sources suggesting that the line
of sight has a considerable inclination to the axis of the jets. The problem arises
{\em only} if the core (pc scale) jets is related to the much larger (100 kpc scale)
jets. There is both theoretical (e. g.  \pcite{valtonen99}) and observational evidence
against this assumption.

\section{Emission from Clouds Illuminated by an Anisotropic Continuum}

Models based on radiative acceleration of optically thick clouds with small volume
filling factor gained  wide acceptance in the Eighties (\pcite{om86} and references
therein; see also \pcite{binette96}). However, problems with cloud confinements and
stability \cite{mathews90} have made them increasingly less frequently invoked to
explain observations.

First RM studies on Balmer lines excluded radial, and favored orbital or chaotic
motions (e.g. \pcite{kg91,kor95}). \scite{goadw96} applied RM techniques to one of the
most extensively monitored objects: NGC 5548. They  ruled out radial motions, and found
that the \civ\ line variations are broadly consistent with a spherical BLR geometry, in
which clouds following randomly inclined circularly Keplerian orbits are illuminated by
an anisotropic source of ionizing continuum. A RM result favoring Keplerian motion may
be approximately correct also for models in which the emitting gas is not bound, such
as a wind, since most of the emission occurs near the base of the flow, when the
velocity is still close to the escape velocity which is similar to the Keplerian
velocity  \cite{mc98}.

\section{Is the disk + wind model applicable to all AGN?}

Emission from a terminal  flow can explain the recent observations of Goad et al. 1999
who reported that LIL (\mgii + \feiiuv) in NGC 3516 do not respond to continuum
variations which did induce detectable variability in the HIL (\lya\ and \civ) lines.
Hydromagnetic wind models such as those developed by \scite{emmering92} and
\scite{bottorff97} exhibit these basic properties.  A two-zone wind provides another
scenario for the different origins of LIL and HIL.

Turning to the general population of RL AGN, the predominance of redshifts and redward
asymmetric profiles is difficult to explain. Several lines of evidence suggest a
significant role of gravitational redshift in RL AGN \cite{corbin97} possibly related
to  a lower distance (in units of gravitational radii) between BLR and central black
hole, which may be systematically more massive in RL than in RQ AGN. If this is the
case, then a double zone wind may be present also in RL AGN, since \civbc\ is still
systematically blueshifted with respect to \hbbc. The ``correlation'' between \civbc\
and \hbbc\ parameters could be due to the impossibility of maintaining a radial flow
along the disk axis, where a relativistic jet is instead propagating. This will make
any HIL outflow possible only at lower latitudes over the disk plane, and therefore
will produce  more similar \hbbc\ and \civbc\ profiles.

\section{CONCLUSION}

While observations support emission from an accretion disk and an associated spherical
wind in RQ AGN with FWHM(\hbbc) $\simlt$ 4000 \kms, there is not enough observational
support to warrant the same conclusion for RL AGN (and possibly RQ with FWHM$>$4000
\kms), although a disk + wind model is a viable possibility also in this case. Wide
separation double peakers (mostly RL) do not provide conclusive evidence in favor of
LIL disk emission; rather, there is evidence against disk emission as well as against
every other reasonably simple scenario.

\acknowledgements DD-H acknowledges support through grant IN109698 from PAPIIT-UNAM. PM
acknowledges financial support from MURST through grant Cofin 98-02-32, as well as
hospitality and support from IA-UNAM. We also acknowledge consistent (cycles5-9 AR and
GO)   rejection of proposals to continue this work.



\end{document}